\documentclass[trackchanges]{aastex63}

\pdfoutput=1

\received{May 03, 2021}
\revised{Jul 01, 2021}
\accepted{Jul 26, 2021}
\submitjournal{ApJ}

\shorttitle{Reconnections in quasi-perpendicular shocks}
\shortauthors{Lu et al.}

\begin{document}

\title{Two-dimensional particle-in-cell simulation of magnetic reconnection in the downstream of a quasi-perpendicular shock}

\correspondingauthor{Quanming Lu, Zhongwei Yang}
\email{qmlu@ustc.edu.cn, zwyang@swl.ac.cn}

\author[0000-0003-3041-2682]{Quanming Lu$^{\dag}$}
\affiliation{CAS Key Lab of Geospace Environment, School of Earth and Space Sciences, University of Science and Technology of China, Hefei 230026, China {\rm{\color{blue}(qmlu@ustc.edu.cn)}}}
\affiliation{CAS Center for Excellence in Comparative Planetology, China}

\author[0000-0002-1509-1529]{Zhongwei Yang$^{\dag}$}
\affiliation{State Key Laboratory of Space Weather, National Space Science Center, Chinese Academy of Sciences, Beijing, 100190, People's Republic of China {\rm{\color{blue}(zwyang@swl.ac.cn)}}}

\author{Huanyu Wang}
\affiliation{CAS Key Lab of Geospace Environment, School of Earth and Space Sciences, University of Science and Technology of China, Hefei 230026, China {\rm{\color{blue}(qmlu@ustc.edu.cn)}}}
\affiliation{CAS Center for Excellence in Comparative Planetology, China}

\author{Rongsheng Wang}
\affiliation{CAS Key Lab of Geospace Environment, School of Earth and Space Sciences, University of Science and Technology of China, Hefei 230026, China {\rm{\color{blue}(qmlu@ustc.edu.cn)}}}
\affiliation{CAS Center for Excellence in Comparative Planetology, China}

\author{Kai Huang}
\affiliation{CAS Key Lab of Geospace Environment, School of Earth and Space Sciences, University of Science and Technology of China, Hefei 230026, China {\rm{\color{blue}(qmlu@ustc.edu.cn)}}}
\affiliation{CAS Center for Excellence in Comparative Planetology, China}

\author[0000-0003-2248-5072]{San Lu}
\affiliation{CAS Key Lab of Geospace Environment, School of Earth and Space Sciences, University of Science and Technology of China, Hefei 230026, China {\rm{\color{blue}(qmlu@ustc.edu.cn)}}}
\affiliation{CAS Center for Excellence in Comparative Planetology, China}

\author{Shui Wang}
\affiliation{CAS Key Lab of Geospace Environment, School of Earth and Space Sciences, University of Science and Technology of China, Hefei 230026, China {\rm{\color{blue}(qmlu@ustc.edu.cn)}}}
\affiliation{CAS Center for Excellence in Comparative Planetology, China}



\begin{abstract}

In this paper, by performing a two-dimensional particle-in-cell simulation, we investigate magnetic reconnection in the downstream of a quasi-perpendicular shock. The shock is nonstationary, and experiences a cyclic reformation. At the beginning of reformation process, the shock front is relatively flat, and part of upstream ions are reflected by the shock front. The reflected ions move upward in the action of Lorentz force, which leads to the upward bending of magnetic field lines at the foot of the shock front, and then a current sheet is formed due to the squeezing of the bending magnetic field lines. The formed current sheet is brought toward the shock front by the solar wind, and the shock front becomes irregular after interacting with the current sheet. Both the current sheet brought by the solar wind and the current sheet associated with the shock front are then fragmented into many small filamentary current sheets. Electron-scale magnetic reconnection may occur in several of these filamentary current sheets when they are convected into the downstream, and magnetic islands are generated. A strong reconnection electric field and energy dissipation are also generated around the X line, and high-speed electron outflow is also formed.

\end{abstract}




\section{Introduction}

Collisionless magnetic reconnection plays an important role in the earth's magnetosphere, and two sites where magnetic reconnection can often be observed are dayside magnetopause and magnetotail. In the dayside magnetopause, magnetic reconnection occurs when the interplanetary magnetic field has a southern component, and plasma energy in the solar wind enters the magnetosphere through magnetic reconnection \citep[e.g.,][]{Dungey1961,Paschmann1979,Pu2007}. Magnetic reconnection in the magnetotail explosively releases the stored magnetic energy in the lobe, and causes substorms \citep[e.g.,][]{Baker2002,Angelopoulos2008,Lu2020b}. Recently, with the availability of high-resolution satellite observations, the transition region or magnetosheath downstream of the bow shock is evidenced to be another site, where reconnection can be often observed in the magnetosphere \citep[e.g.,][]{Retino2007,Yordanova2016,Voros2017,Phan2018,Wang2018}.

The bow shock is formed after the high-speed solar wind interacts with the earth's magnetosphere, and the magnetosheath behind the shock is usually in a turbulent state. According to the shock normal angle ($\theta_{Bn}$) between the shock normal and the upstream magnetic field, the bow shock can be separated into a quasi-parallel shock ($\theta_{Bn}<45^{\circ}$) and quasi-perpendicular shock ($\theta_{Bn}>45^{\circ}$), and their characteristics are quite different. In a quasi-parallel shock, the reflected ions by the shock can move into the far upstream and excited low-frequency large-amplitude electromagnetic waves due to plasma beam instabilities \citep[e.g.,][]{Hao2016,Scholer1990,Su2012}. With a global three-dimensional (3-D) hybrid simulation model, \citet{Lu2020a} found that when these waves are convected toward the shock by the solar wind, they evolve into current sheets after penetrating the shock. Magnetic reconnection can occur in these current sheets and magnetic islands may be generated. By performing a local two-dimensional (2-D) particle-in-cell (PIC) simulation, \citet{Bessho2019} found that magnetic reconnection can also occur in the transition region of a quasi-parallel shock. In situ evidences of magnetic reconnection in the magnetoshearth downstream of a quasi-parallel shock have also been provided by Cluster and MMS satellite observations \citep[e.g.,][]{Phan2018,Retino2007,Voros2017,Yordanova2016}.

In a perpendicular shock, the reflected ions by the shock quickly transmit to the downstream, where they have an anisotropic distribution with the perpendicular temperature larger than the parallel temperature, and then excite ion cyclotron waves and mirror waves \citep[e.g.,][]{Hao2014,Lu&Wang2006,McKean1992,Winske1988}. Although there are some evidences indicating the existence of magnetic reconnection in the transition region of a quasi-perpendicular shock \citep{Wang2019}. In this paper, with the help of a 2-D PIC simulation model, we try to figure out the mechanism for the formation of current sheets and consequential magnetic reconnection in a quasi-perpendicular shock.

\section{Simulation model}

An open-source electromagnetic 2-D PIC simulation code named EPOCH \citep{Arber2015} are used in this letter to study magnetic reconnection in a perpendicular shock. The shock is formed by the injection method \citep{Matsukiyo&Scholer2012}, where particles are injected from the left side of the simulation boundary (at $x=0$) with a speed $V_{in}=7V_{A0}$ (where $V_{A0}$ is the Alfv\'{e}n speed based on the upstream plasma density $n_0$ and magnetic field $B_0$) and a specular reflection for particles is used in the right boundary ($x=L_x$, where $L_x$ is the size of the simulation domain in the $x$ direction). The formed shock propagates toward the left (the $-x$ direction). The 2D simulation is performed in the $x-z$ plane, and the ambient magnetic field ${\bf B}_0=B_0(cos\theta_{Bn}{\bf i}_x-sin\theta_{Bn}{\bf i}_y)$, where $\theta_{Bn}$ is the shock normal angle. In our simulation, we choose $\theta_{Bn}=600$, therefore, the shock is quasi-perpendicular. Here, the ambient magnetic field have a strong component perpendicular to the simulation plane, as has been done in \citep{Yang2010,Yang2012}. Periodic boundary conditions for both electromagnetic fields and particles are applied in the $y$ direction.
The domain size of the simulation is $L_x\times L_y=470.4d_{i0}\times12d_{i0}$, where $d_{i0}=c/\omega_{pi,0}$ is ion inertial length and $\omega_{pi,0}=\sqrt{n_0e^2/m_i\varepsilon_0}$ is the ion plasma frequency based on the upstream plasma density $n_0$. The grid number is $n_x\times n_y=47040\times1200$, and grid size is $\Delta x=\Delta y=0.01d_{i0}$. Initially, there are $50$ ions and electrons in every cell, and the ion-to-electron mass ratio is $m_i/m_e=64$. The light speed is $c/V_{A0}=28$. The plasma beta values in the upstream are $\beta_{i0}=0.1$, $\beta_{e0}=0.2$.

\section{Simulation results}

The shock is formed due to the interaction of the injected plasma from the left boundary and the plasma reflected from the right boundary, and its front propagates from right to left. Figure 1 describes the evolution of the perpendicular shock by plotting the stacked profiles of the magnetic field $\overline{B}_t$ from $\Omega_{i0}=5$ to $20$. During this time period, the shock front is already sufficiently away from the right boundary, and influence from the right boundary can be negligible. Here, $\overline{B}_t$ is the average value of the magnetic field $B_t=\sqrt{B_x^2+B_y^2+B_z^2}$ over the z direction. The propagation speed of the shock front is about $2.5V_{A0}$, and then the Alfv\'{e}n Mach number is about 9.5 in the shock frame. The reformation of the shock front can be clearly identified in the figure, which is demonstrated by the cyclic variations of the magnetic field $\overline{B}_t$ in the shock front. The reformation period is estimated to be about $2.1\Omega_{i0}$. It is generally considered that the reformation of the shock front is related to the reflected ions by the shock \citep[e.g.,][]{Lembege&Savoini2002,Yang2009,Yang2020b}. The reflected ions are accumulated in the foot, and then the foot amplitude increases until exceeds that of the shock front. At last, a new shock front is formed, and the old shock front become weaker and weaker.

The reformation process of the shock front can be exhibited more clearly in Figure 2, which shows the magnetic field lines and electron current in the $y$ direction ($j_{ey}$) at $\Omega_{i0}t=9.3$, 9.7, 9.9, 10.7, 11.1 and 11.4. The profiles of the average value of the magnetic field $\overline{B}_t$ at these times are also plotted in the figure for reference. These time slots are selected from one reformation period (from about  9.3 to 11.4). At $\Omega_{i0}t=9.3$, the shock front is relatively flat, and it is located at $x\approx446d_{i0}$. There exists strong electron currents associated with the shock front. Part of upstream ions are reflected by the shock front, and they will move upward in the action of the Lorentz force, which drags the magnetic field lines at the foot region to bend upward. Then, the bending magnetic field lines are squeezed. At last, a current sheet is formed, where the current is carried mainly by the electrons and points to the $+y$ direction. When the current sheet is convected toward to the shock front by the solar wind. The interaction of these structures with the shock front causes local curvature variations in the shock front, and a rippled shock front is formed. From the profile of the average value of the magnetic field $\overline{B}_t$, we can identify the enhancement of the magnetic field in the foot region of the shock. Simultaneously, both the current sheet brought by the solar wind and the current sheet associated with the shock front are distorted, and several magnetic islands are generated in the foot region and downstream. When these structures are merged totally with the shock front, the enhancement of the magnetic field surpass that of the shock front, and a new shock front is at last formed. A new cycle of shock reformation will begin again.

The generation of magnetic islands in the distorted currents sheets in the foot region and downstream of the shock indicates the occurrence of magnetic reconnection. This can be demonstrated more clearly in Figure 3. In Figure 3, we plot the enlarged view of the denoted region by the red lines in Figure 2(c). Figure 3(a)-(d) present the electron current in the $y$ direction $j_{ey}$, the fluctuating magnetic field in the $y$ direction $\delta B_y$ (where $\delta B_y=B_y-\widetilde{B}_y$, and $\widetilde{B}_y$ is the average value of $B_y$ in the denoted region), the electric field in the $y$ direction $E_y^{'}$ (where $\bf{E^{\prime}}=\bf{E}+\bf{V_e}\times\bf{B}$), and $j_{ey}E_y^{\prime}$ at $\Omega_{i0}t=9.9$, respectively. The magnetic field lines are also plotted in these figures for reference. There are several strong electron current sheet in the region, and we focus on two strong electron current sheet denoted by CS1 and CS2. We have also changed the two current sheets into the local current sheet coordinate systems $(L_1,M_1,N_1)$ and $(L_2,M_2,N_2)$ with the Minimum Variance Analysis \citep{Paschmann1998}, and $\bf{L}_1=$(0.585, 0, 0.81) and $\bf{L}_2=$(0.842, 0, 0.538). The currents are directed toward the $+y$ and $-y$ directions in the current sheets CS1 and CS2 respectively, and an X type configuration of magnetic field lines is formed in every current sheet. Their half-widths of the current sheets are estimated to be about $0.19d_{i0}$ and $0.11d_{i0}$  around the X lines, while their lengths are about $0.28d_{i0}$ and $0.36d_{i0}$. The fluctuating magnetic field $\delta B_y$ is produced around the two X lines. Here, the fluctuating magnetic field $\delta B_y$ does not exhibit the well-known quadrupolar structure as has been found in other simulations and satellite observations in the Harris current sheet \citep[e.g.,][]{Fu2006,Huang2010,Wang2012}. This may be caused due to the curvature of magnetic field lines around the X line and fast evolution of the current sheets. The reconnection electric field $E_y^{'}$ as well as $j_{ey}E_y^{'}$ exist in the vicinities of two line X lines. The peak values of the reconnection electric field are about $2.6V_{A0}B_0$ and $-2.3V_{A0}B_0$ in the current sheets CS1 and CS2, respectively. In Figure 3(e) and (f), we show that the electron outflows along the $L_1$ and $L_1$ directions in the two current sheets. Here, the background electron flow in the two current sheets have been eliminated. Obviously, the two X lines are accompanied by the bi-directional high-speed electron outflows, and their speed may exceed the local Alfv\'{e}n speed (here, values of the local Alfv\'{e}n speed are about $1.9V_{A0}$ and $1.8V_{A0}$, respectively).

Now we analyze in detail the characteristics of the current sheet CS1, which is asymmetric. In the current sheet, the magnetic field and plasma density at the right side are $B_1\approx4.26B_0$ and $n_1\approx4.02n_0$, while those at the left side are $B_2\approx4.69B_0$ and $n_2\approx4.47n_0$. Therefore, the half width and length of the current sheet are about 0.38 and 0.56 local ion inertial length, where the local ion inertial length is calculated based on the density $n_m=(n_1B_2+n_2B_1)/(B_1+B_2)\approx4n_0$. Equivalently, the half width of the current sheet is about 3.04 local electron inertial length. The reconnection electric field normalized by $V_AB_m$ is about 0.26, where $B_m=2B_1B_2/(B_1+B_2)\approx4.5B_0$ and $V_A^2=B_1B_2/(\mu_0m_in_m)\approx5.0$ \citep{Cassak&Shay2007}. According to the results in \citet{Pyakurel2019}, the ions do not response to magnetic reconnection when the length of the current sheet is comparable to the ion inertial length. Simultaneously, we can also find that the life time of the current sheet is shorter than the ion cyclotron period based on $B_m(2\pi/\Omega_{ci}\approx1.39)$ as shown in Figure 4. Based on these considerations, only the electrons are involved in reconnection, and it is electron magnetic reconnection without ion coupling. Therefore, the reconnection electric field normalized by $V_{Ae}B_m$ (where $V_{Ae}=B_1B_2/(\mu_0m_en_m)\approx320$) is about 0.01.

The following evolution of the current sheets in the denoted region by the red lines in Figure 2(c) is also investigated. Figure 4 plots the electron current in the $y$ direction $j_{ey}$ and the topology of the magnetic field lines at $\Omega_{i0}t=9.9$, 10.1, 10.3, 10.5, 10.7, and 11.1. We still focus on the two current sheets denoted by CS1 and CS2. In each current sheet, two magnetic islands are formed around the X line, and they become larger and larger with the proceeding of magnetic reconnection. The sizes of these islands ranges from $1.2d_{i0}$ to $1.57d_{i0}$, or 1.74 to 2.26 local ion inertial lengths. The islands may also merge with other islands. Simultaneously, the two current sheets are distorted largely, or even broken into several segments.

\section{Conclusions and discussions}

In this paper, with a 2-D PIC simulation model, we have studied magnetic reconnection associated with a quasi-perpendicular shock. The results showed that the quasi-perpendicular shock is non-stationary, and it experiences a reformation with a cyclic period about 2.1$\Omega_{i0}^{-1}$. During the reformation of the shock, the magnetic field lines in the foot region are firstly bent upward and then squeezed. A current sheet is at last formed. Because the width of the current sheet is on the electron scale, the electron motions are non-adiabatic, and the current is carried mainly the electrons. The current sheet is convected to the shock front by the solar wind. After these structures interact with the shock front, the shock front become irregular. Both the current sheet brought by the solar wind and the current sheet associated with the shock front are distorted and fragmented into many filamentary current sheets in the shock transition region. Magnetic reconnection may occur in these filamentary current sheets, and magnetic islands are formed. Based on the width, length and life time of these current sheets, we can know that only electrons are involved in magnetic reconnection.

Electron magnetic reconnection without ion coupling have been observed by MMS satellites in the magnetotail \citep{Wang2018} and magnetosheath \citep{Phan2018}. The observations by \citet{Phan2018} have indicated that electron reconnection can occur in the downstream of a quasi-parallel shock. However, our results predict that electron magnetic reconnection may also occur in the transition region and downstream of a quasi-perpendicular shock, which is waiting for the verification by MMS observations in the future. Please note, in this paper, the simulation is performed in the $x-z$ plane, and the ambient magnetic field have a strong $y$ component. Therefore, magnetic reconnection in the shock transition region of the perpendicular shock has a strong guide field. How it will influence the energy dissipation in a quasi-perpendicular shock is our further investigation.

\acknowledgments

We are grateful to the valuable suggestions with C. Huang from IGG. This work was supported by the NSFC Grants 41774169, 41804159, Key Research Program of Frontier Sciences, CAS (QYZDJ-SSWDQC010), and the Strategic Priority Research Program of Chinese Academy of Sciences, Grant No. XDB 41000000. Z. Y. acknowledges partial support by the Specialized Research Fund for State Key Laboratories of China, Youth Innovation Promotion Association of the CAS (2017188), and the project of Civil Aerospace ``13th Five Year Plan" Preliminary Research in Space Science (Project Name: Research on Important Scientific Issues of Heliospheric Boundary Exploration, Project No.: D020301£¬D030202). The computations are performed by Numerical Forecast Modeling R\&D and VR System of State Key Laboratory of Space Weather, and HPC of Chinese Meridian Project.\\

%



\bibliography{references}{}

\clearpage

\begin{figure}
\epsscale{1.0}
\plotone{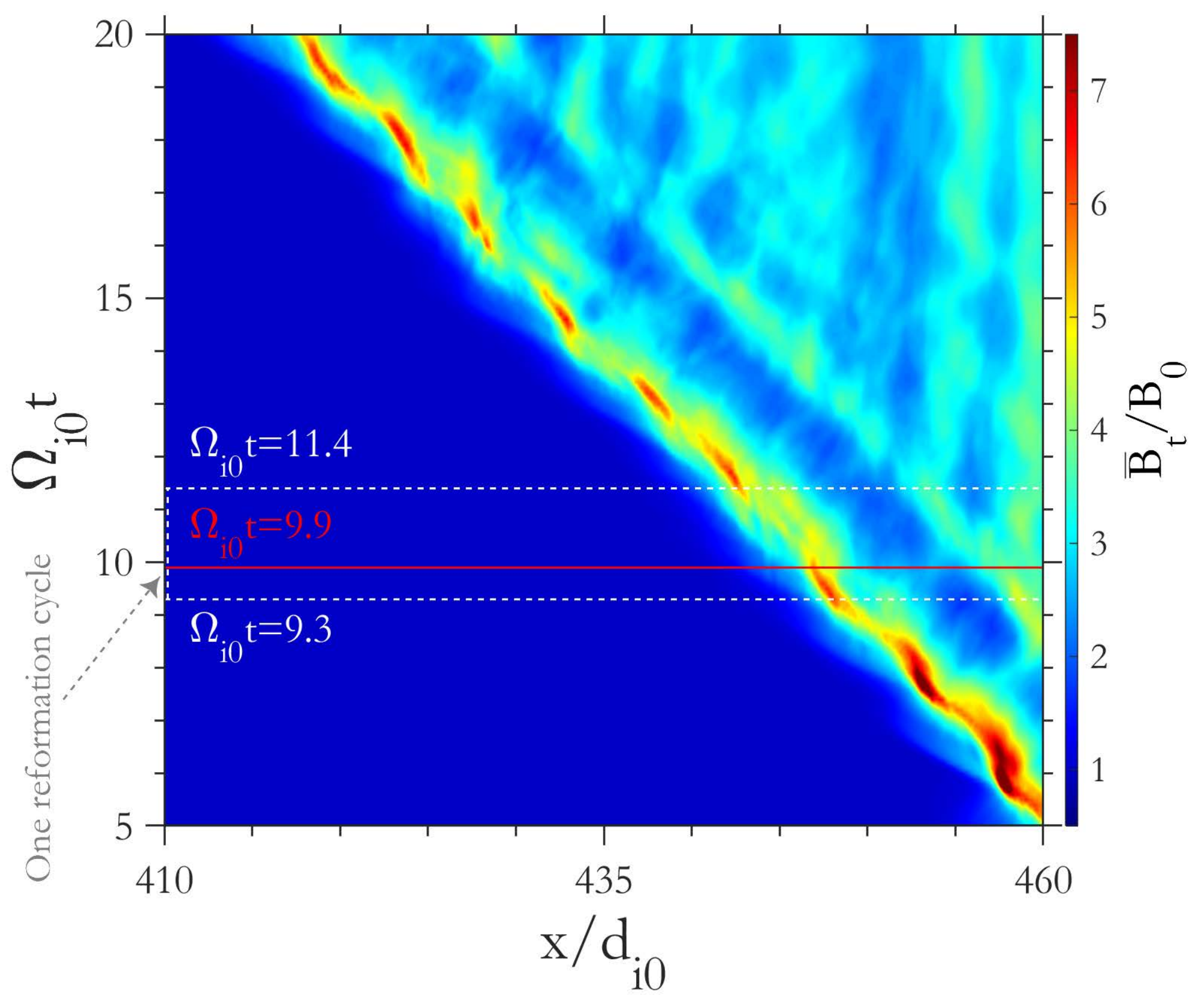}
\caption{\label{fig:fig1}The evolution of the perpendicular shock by plotting the stacked profiles of the magnetic field $\overline{B}_t$ from $\Omega_{i0}t=5$ to 20. Here, $\overline{B}_t$ is the average value of the magnetic field $\overline{B}_t=\sqrt{B_x^2+B_y^2+B_z^2}$ over the $z$ direction.}
\end{figure}

\clearpage

\begin{figure}
\epsscale{1.1}
\plotone{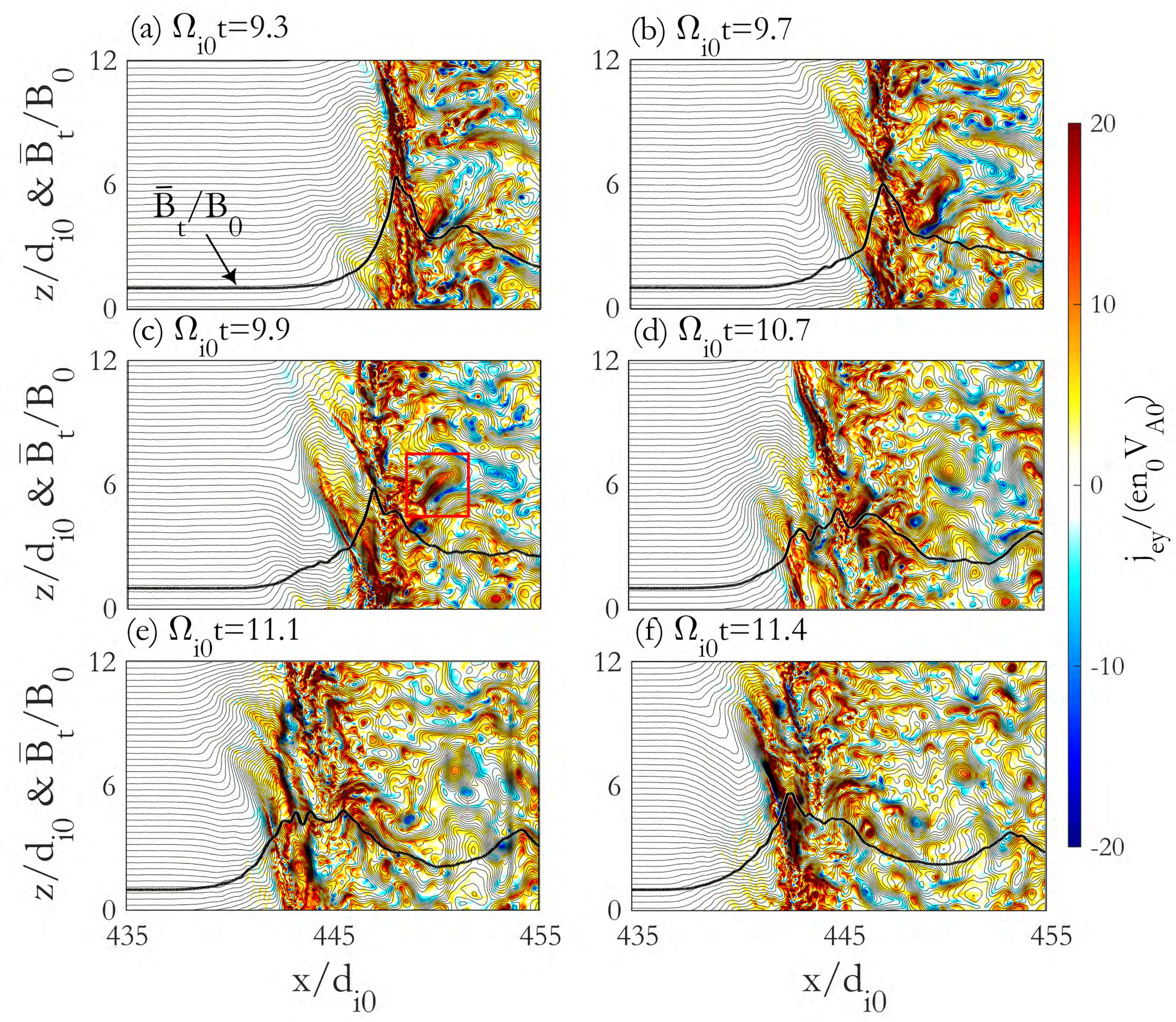}
\caption{\label{fig:fig2}The magnetic field lines and electron current in the $y$ direction ($j_{ey}$) at $\Omega_{i0}t=$(a)9.3, (b)9.7,(c) 9.9, (d)10.7, (e)11.1 and (f)11.4. The profiles of the average value of the magnetic field $\overline{B}_t$ are also plotted in the figure for reference.}
\end{figure}

\clearpage

\begin{figure}
\epsscale{1.1}
\plotone{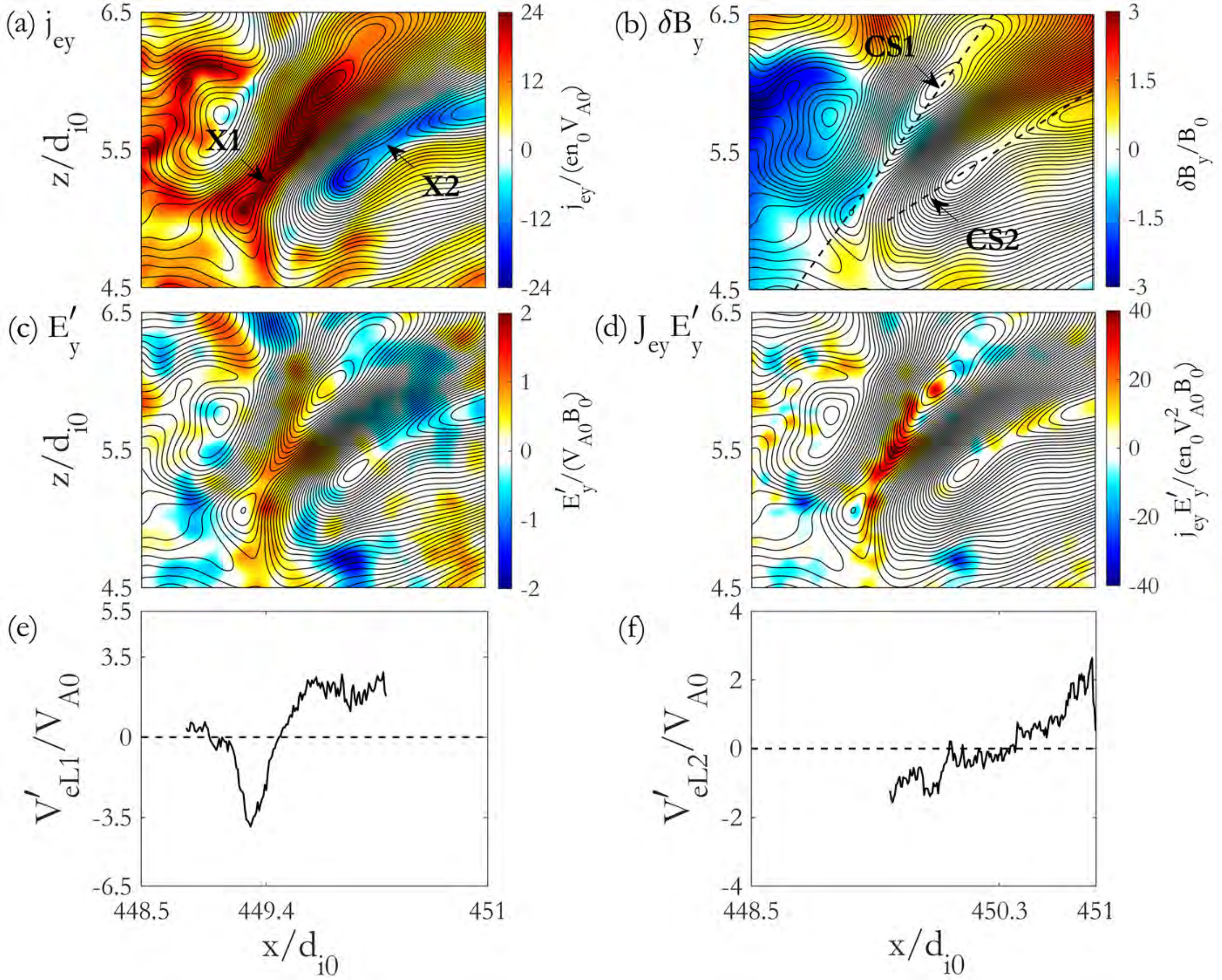}
\caption{\label{fig:fig3}The enlarged view of the denoted region by the red lines in Figure 2(c). (a) the electron current in the $y$ direction $j_{ey}$, (b) the fluctuating magnetic field in the $y$ direction $\delta{B}_y$ (where $\delta B_y=B_y-\widetilde{B}_y$, and $\widetilde{B}_y$ is the average value of $B_y$ in the denoted region), (c) the electric field in the $y$ direction $E_y^{'}$ (where ${\bf E}^{'}={\bf E}+{\bf V}_e\times{\bf B}$), (d) $j_{ey}E_y^{'}$ at $\Omega_{i0}t=9.9$. In the figure, CS1 and CS2 denote two current sheets. The X1 and X2 denote the X points in the current sheets CS1 and CS2. (e)¨C(f) show that the electron outflows along the $L_1$ and $L_2$ directions in the two current sheets. Here, the background electron flow in the two current sheets have been eliminated.}
\end{figure}

\clearpage

\begin{figure}
\epsscale{1.2}
\plotone{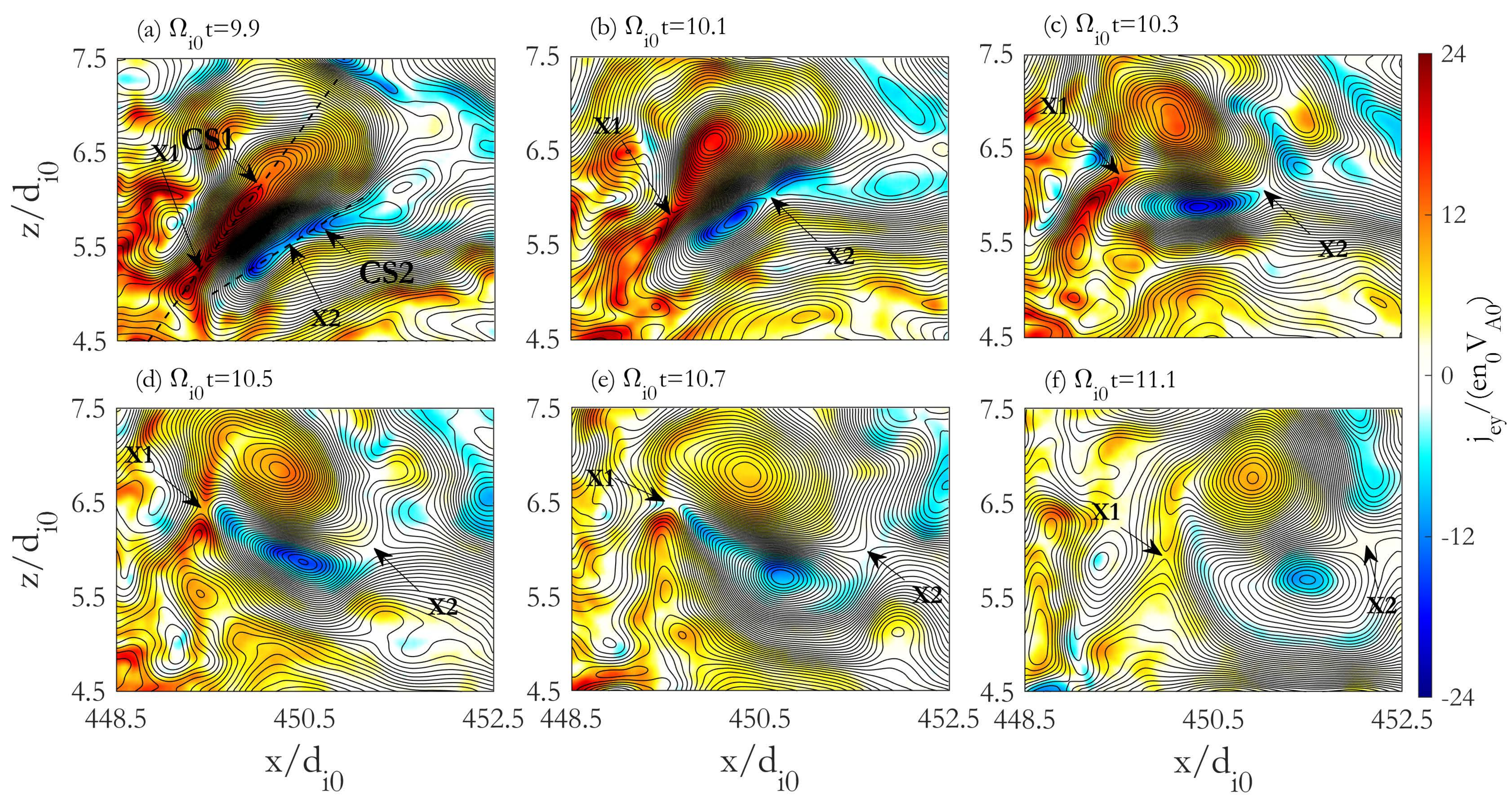}
\caption{\label{fig:fig4}The electron current in the $y$ direction $j_{ey}$ and the topology of the magnetic field lines at $\Omega_{i0}t=$(a)9.9, (b)10.1, (c)10.3, (d)10.5, (e)10.7 and (f)11.1. The X1 and X2 in the figure stand for the X points in the current sheets CS1 and CS2, which is shown in Figure 4(a).}
\end{figure}

\clearpage



\end{document}